\documentclass[12pt]{article}
\usepackage{graphics}
\begin{document}
\markright{Gravitons ...}
\title{Aspects of Graviton Detection: Graviton Emission and Absorption by Atomic Hydrogen}

\author{Stephen Boughn$^*$ and Tony Rothman$^\dagger$
\\[2mm]
{\small\it \thanks{sboughn@haverford.edu}}~ \it Haverford College , \\
\it Haverford PA, 09140\\
{\small\it\thanks{trothman@princeton.edu}}~ \it Princeton University, \\
\it Princeton, NJ 08544 }

\date{{\small   \LaTeX-ed \today}}

\maketitle

\begin{abstract}
 Graviton absorption cross sections and emission rates for hydrogen are
calculated by  both semi-classical and field theoretic methods. We
point out several mistakes in the literature concerning
spontaneous emission of gravitons and related phenomena, some of
which are due to a subtle issue concerning gauge invariance of the
linearized interaction Hamiltonian.

\vspace*{5mm} \noindent PACS: 03.65.Sq, 04.30.Db, 04.30.Nk,
04.60.-m, 95.30.Cq, 95.30.Sf
\\ Keywords: Gravitons, Gravitational Waves, Graviton Emission, Graviton Absorption

\end{abstract}
\section{Introduction}
\label{intro}

Two years ago, Dyson  published a conjecture that no
conceivable experiment performed in our universe could detect a
single graviton \cite{Dyson04}. Recently, in a  companion paper
\cite{RB06}(henceforth RB), we have addressed Dyson's proposition
and have been unable to find any clear-cut fundamental reason
forbidding the detection of one graviton. However, when anything
remotely resembling ``real" physics is taken into account, a
detection becomes impossible, making Dyson's conjecture very
likely true, at least without the introduction of exotic physics,
such as extra dimensions.

In the process of checking our calculations against graviton
spontaneous emission rates for hydrogen, we have discovered
several mistakes in the literature involving both spontaneous
emission rates and the response of elastic media to classical
gravitational waves. Although some of the mistakes are evidently
numerical, others involve a rather subtle issue of gauge
invariance of the interaction Hamiltonian in linearized gravity.
Our purpose here is to present these calculations in detail,
thereby illuminating the source of the difficulties. To
demonstrate the validity of our results, we compute transition
rates from the 3d to 1s hydrogen states using both semi-classical
and field theoretic methods in two different gauges.  We also
present the consistency check that led to the discovery of the
mistakes: the requirement that detailed balance be satisfied among
the spontaneous emission, stimulated emission and absorption
rates. Finally, we compute the graviton ionization cross section
of hydrogen, because in RB we chose this gravitational analogy to
the ordinary photoelectric effect as the most relevant one in
addressing Dyson's question.

\section{Semi-classical Analysis of Spontaneous Graviton Emission}
\setcounter{equation}{0} \label{spont}

In the standard semi-classical treatment of spontaneous emission
of photons from atomic dipole transitions (see, for example,
Schiff \cite{Schiff49}), the transition rate is found by beginning
with the classical expression for power emitted by a dipole,
\begin{eqnarray}
P = \frac{4k^2}{3c}{| {\bf J}_0 |}^2, \label{epower}
\end{eqnarray}
in which ${\bf J}_0 \equiv \int {\bf J}({\bf r})d^3 r$ is the
total current and $k$ is the wave number of the emitted radiation.
One then attempts to interpret this expression in a quantum
mechanical way by identifying the current density ${\bf J}$
(charge density times velocity) with the probability current $e{|
\Psi |}^2 {\bf p}/m$, for wavefuction $\Psi$ and electron charge
$e$. Because the quantum mechanical transition is between states
two states $a$ and $b$, the standard treatment replaces $e{| \Psi
|}^2$ with $e\Psi_a^* \Psi_b$, and so the total
 current  becomes ${\bf J}_0 = \frac{e}{m} \int
\Psi_a^* {\bf p} \Psi_b d^3 r$. Finally, the quantum mechanical
transition rate is assumed to be
\[
\Gamma = \frac{P}{\hbar\omega},
 \]
with $P$ given by inserting the ``quantum" expression for ${\bf
J}_0$ into Eq. (\ref{epower}). The justification for this
 argument is, to say the least, tenuous; therefore it is
important to verify the resulting transition rate in some other
way, e.g., that it satisfies detailed balance or that a ``proper''
field theoretic calculation yields the same result
\cite{Schiff49}.

For the case of spontaneous graviton emission, we begin with the
(quadrupole) gravitational analog of Eq. (\ref{epower}),
\begin{eqnarray}
P = \frac{2G\omega^6}{5c^5} \langle I_{jk} I_{jk}\rangle.
\label{gpower}
\end{eqnarray}
In this expression, $I_{jk} \equiv \int \rho({\bf r})(x_jx_k -
\frac{1}{3} \delta_{jk}r^2)d^3r$ is the reduced mass quadrupole
moment, ${\bf r} = \{x_j\}$ are cartesian coordinates in the local
inertial frame, and $\rho$ is the mass density of the radiating
system \cite{Misner73}. The normalization for $I_{jk}$, as well as
${\bf J}$ above, is usually taken to be $I_{jk}({\bf r},t) =
I_{jk}({\bf r})e^{-i\omega t} + c.c.$. Following the
electromagnetic procedure, the  semi-classical approach for
gravity substitutes the electron mass $m_e$ for $e$ and replaces
$\rho$ by $m_e \Psi_{b}^*\Psi_{a}$. (The equivalent mass density
associated with the electric fields of the electron and proton is,
for non-relativistic systems, small compared to the electron mass;
we discuss this issue in more detail in \S\ref{hamiltonian}.)
Therefore the reduced quadrupole moment becomes
\begin{eqnarray}
I_{jk} = m_e \int \Psi_{b}^*(x_jx_k - \frac{1}{3}
\delta_{jk}r^2)\Psi_{a} d^3r . \label{quad}
\end{eqnarray}
To compare with previous work, we can now compute the transition
rate from the $3d2$ to the $1s$ state of hydrogen.  In terms of
the Bohr radius $a = \hbar^2/m_e e^2$, the normalized
wavefunctions are
\begin{eqnarray}
\Psi_{1s} = \frac{1}{\sqrt{\pi} a^{3/2}}e^{-r/a}  &;&
\Psi_{3d2}=\frac{1}{162\sqrt \pi}
\frac{1}{a^{3/2}}\left(\frac{r^2}{a^2}\right)e^{-r/3a}sin^2\theta
e^{2i\phi}.\label{wavefcns1}
\end{eqnarray}
 Inserting these
expressions into into Eq. (\ref{quad}) and subsequently into Eq.
(\ref{gpower}) yields a transition rate of
\begin{eqnarray}
\Gamma = P/\hbar\omega = \frac{3^8 Gm_e^2 a^4 \omega^5}{5 \times
2^{13}\hbar c^5} = \frac{\alpha^6G{m_e}^3c}{360\hbar^2}
 = 5.7 \times 10^{-40} s^{-1} \label{Gamma}
\end{eqnarray}
where $\alpha = e^2/\hbar c$ is the fine structure constant. It is
straightforward to show that the transition rates for all $3d$
states ($3d(\pm2)$, $3d(\pm1)$, and $3d0$) are all the same.

In his standard text {\it Gravitation and Cosmology}
\cite{Weinberg72}, Weinberg presented a calculation of this
transition rate using the same semi-classical approach. His result
was $\Gamma = 2.5 \times 10^{-44} s^{-1}$, which differs from ours
by more than four orders of magnitude. Since the methods are
identical, and the dimensional factors agree, the discrepancy is
presumably due to a numerical error.

Indeed, in his book on quantum gravity \cite{Kiefer04}, Kiefer
obtained the result in Eq. (\ref{Gamma}), again using a
semiclassical analysis.  While this agreement may indicate the
above analysis has no computational errors, the  semi-classical
treatment itself ``can claim only a moderate amount of
plausibility'' \cite{Schiff49}.  Stronger support is afforded by
demonstrating that the transition rate of Eq. (\ref{Gamma})
satisfies detailed balance. To do this it is first necessary to
evaluate the cross sections for absorption and stimulated
emission.

\section{Absorption, Stimulated Emission, and Detailed Balance}
\setcounter{equation}{0} \label{balance}

Unlike spontaneous emission, the processes of absorption and
stimulated emission can be treated adequately in the context of
classical quantum mechanics, that is, without the use of quantum
field theory.  According to first order perturbation theory, the
transition probability between two hydrogenic states $\Psi_a$ and
$\Psi_b$ is proportional to the square of the matrix element
$\langle \Psi_b | H | \Psi_a \rangle$, where the interaction
Hamiltonian $H$ is derived from the interaction Lagrangian by its
definition $H = pv - L$.  For metric deviation $h_{\mu\nu} =
g_{\mu\nu} - \eta_{\mu\nu} < < 1$ and stress-energy tensor
$T^{\mu\nu}$, the interaction Lagrangian density is given by
\cite{Weinberg72,Dyson69}
\begin{eqnarray}
\mathcal{L} = \frac{1}{2} h_{\mu\nu}T^{\mu\nu}. \label{interham}
\end{eqnarray}
Although the action from which $\cal L$ is derived is a scalar,
$\cal L$ itself is not gauge invariant.  It is in fact this
property that has led to the more serious conceptual errors
mentioned in the Introduction and which we will discuss in detail
in \S\ref{hamiltonian}. The standard choice is to work in a {\em
local inertial frame} (LIF) in which case the dominant term of the
stress-energy tensor is  just the mass-energy density, and so
$\mathcal{L} \approx \frac{1}{2} h_{00}T^{00}$.  When the
generalized velocities are negligible, as in an LIF, the interaction
Hamiltonian density is simply $\mathcal{H} = - \mathcal{L}$, and
the  corresponding interaction Hamiltonian for a localized mass
$m$ is therefore $H \approx -\frac{1}{2}mh_{00}$, if we assume
that $h_{00}({\bf r})$ is approximately constant in the region of
nonvanishing mass density.  Although LIFs are the most nearly
Minkowskian and hence physical frame of reference, it is in the
transverse-traceless, or TT, gauge that gravitational waves are
most easily interpreted.

The TT gauge is a subset of what is variously called the Hilbert,
Einstein, Fock, de Donder or Harmonic gauge, which is analogous to
the Lorentz gauge of electromagnetism and  defined by the
requirement that $h^\mu\,_{\nu,\mu} = \frac{1}{2} h
^\mu\,_{\mu,\nu}$. The TT gauge makes the further choice that
$h_{\mu 0} = h^\mu\, _\mu = 0$.  Then for an amplitude $h$ and
polarization tensor $e_{\mu\nu}$ the metric deviation can be
written as $h_{\mu\nu} = h e_{\mu\nu}$, where
$e _{00} = e_{\mu 0} = e^\mu\, _\mu = 0$.  Consequently, a
harmonic, plane gravitational wave (GW) can  be expressed as
\begin{eqnarray}
h^{TT}_{jk} = h e^{i(k_lx^l-\omega t)} e_{jk} + c.c. \label{norm}
\end{eqnarray}
where $j,k,l = 1,2,3$ and $\omega = kc$.
 In this case the Hilbert condition becomes $e_{kl} k^l = 0$, which for a plane wave propagating in
the  z-direction gives $e_{xx}= -e_{yy}= e_{xy}=e_{yx}$
 and $e_{zj} = 0$. In what follows, we normalize the nonzero components of the polarization tensor
 to $|e_{ij}| = {1}/{\sqrt{2}}$.

Independent of normalization,  $h_{00}$ in a LIF can be expressed in terms of 
the quantities $h^{TT}_{jk}$ of the TT gauge.  
Since the duration of the interaction of the incident GW with the 
physical system (the hydrogen atom in this case) may be long, it is necessary to 
use a coordinate system that is locally intertial for an extended time.  There are 
an infinite number of such coordinate systems; however, it can be shown that to 2nd 
order in coordinate displacement $h_{00}$ is the same in all such systems.  Perhaps 
the most useful of these are ``Fermi normal coordinates'' in which
\cite{Misner73}
\begin{eqnarray}
h_{00} = - R_{0j0k} x^j x^ k \label{fermi}
\end{eqnarray}
where $R_{\alpha\beta\gamma\delta}$ is the Riemann curvature tensor.  To lowest order in
$h_{\alpha \beta}$
\begin{eqnarray}
R_{\alpha\beta\gamma\delta} = \frac{1}{2}(h_{\alpha\delta,\beta\gamma}
+ h_{\beta\gamma,\alpha\delta} -  h_{\beta\delta,\alpha\gamma} -  h_{\alpha\gamma,\beta\delta})
\label{riemann}
\end{eqnarray}
and is gauge invariant.
Combining Eqs. (\ref{norm}), (\ref{fermi}), and (\ref{riemann}) yields
\begin{eqnarray}
h_{00} = -\frac{1}{2} \omega^2 he^{i(k_lx^l-\omega t)} x^j x^ k
e_{jk} + c.c. \nonumber \label{h00}
\end{eqnarray}
The LIF interaction Hamiltonian therefore becomes
\begin{equation}
H = \frac{1}{4}m_e \omega^2 h x^j x^k e_{jk} e^{i(k_lx^l-\omega
t)} + c.c..\label{hamiltonian1}
\end{equation}
 This is the Hamiltonian we will use in computing the matrix elements.\\

To calculate the transition rate, one also needs the flux of
incident GWs, which is given by \cite{Misner73}
\begin{equation}
\mathcal{F} = \frac{c^3}{32 \pi G} \langle h^{TT}_{jk,0}
h^{TT}_{jk,0} \rangle, \label{flux}
\end{equation}
where $\langle \rangle$ indicates an average over several cycles.
Assuming equal amplitudes for the two polarizations and the
normalization $|e_{ij}| = 1/{\sqrt{2}}$, we have
\begin{eqnarray}
\mathcal{F} = \frac{c^3 \omega^2}{8 \pi G} h^2.
\nonumber
\end{eqnarray}
The transition probability per unit time between two discrete
states is not constant in time if the incident radiation field and
Hamiltonian are strictly monochromatic.  So we assume that the
radiation is spread over a range of frequencies with uncorrelated
phases. In a small neighborhood $\Delta \omega$ of each of these
frequencies the flux is given by $d\mathcal{F} = I(\omega) \Delta
\omega$ where $I(\omega)$ is the GW intensity.  Therefore,
\begin{eqnarray}
h^2 = \sum_\omega \frac{8 \pi G}{c^3 \omega^2} I(\omega) \Delta
\omega .\nonumber
\end{eqnarray}
 The transition probability
per unit time is then given by standard perturbation theory
\cite{Schiff49}:
\begin{eqnarray}
\Gamma &= &\frac{1}{t} \sum_\omega \frac{4 {| \langle \Psi_m | H |
\Psi_n \rangle |}^2
\sin^2\frac{1}{2}(\omega_{mn}-\omega)t}{\hbar^2
(\omega_{mn}-\omega)^2} \nonumber \\
&&\nonumber\\
& = & \frac{2 \pi G m_e^2}{\hbar^2 c^3 t}\, |\int\Psi_m^*
e^{ik_lx^l}x^jx^ke_{jk}\Psi_n d^3r|^2\, \sum_\omega \frac{\omega^2
I(\omega) \Delta \omega \sin^2\frac{1}{2}(\omega_{mn}-\omega)t}
{(\omega_{mn}-\omega)^2}.\nonumber\\
 \label{abs}
\end{eqnarray}
Here, the difference between the two electron energy levels is
$\Delta E = \hbar \omega_{mn}$ and the sum is over the frequencies
of the phase independent GWs each within a band $\Delta \omega$.
In the limit that $\Delta \omega$ is infinitesimally small, the
summation can be replaced by an integral.  Since the time factor has a
sharp maximum at $\omega = \omega_{mn}$, $I(\omega)/\omega^2$ can
be taken outside the integral and the limits extended to $\pm
\infty$.  Eq. (\ref{abs}) then reduces to
 \vspace{2 mm}
\begin{eqnarray}
\Gamma = \frac{\pi^2 Gm_e^2 \omega_{mn}^2I(\omega_{mn})}{\hbar^2
c^3} |\int\Psi_m^* e^{ik_lx^l}x^jx^ke_jk\Psi_nd^3r|^2.\nonumber
\label{abs2}
\end{eqnarray}

 Finally,  for the transition from the 3d to 1s states of
hydrogen, the wavelength of the GW is much larger than the extent
of the wavefunction so that $e^{ik_lx^l} \approx 1$ (the ``dipole"
approximation). Then
\begin{eqnarray}
\Gamma &= &\frac{\pi^2 Gm_e^2 \omega_{mn}^2
I(\omega_{mn})}{\hbar^2
c^3} |\int\Psi_m^* x^jx^ke_{jk}\Psi_nd^3r|^2\nonumber\\
&&\nonumber\\
 & =& \frac{\pi^2 Gm_e^2
\omega_{mn}^2I(\omega_{mn})}{\hbar^2 c^3}|D_{jk}e_{jk}|^2,
\label{abs3}
\end{eqnarray}
where
\begin{eqnarray}
D_{ij} \equiv \int\Psi_b x_i x_j\Psi_a d^3r. \label{Dqm}
\end{eqnarray}

For a GW  propagating in the $z$ direction, the two linear
polarizations are as given above and
\begin{eqnarray}
|D_{jk}e_{jk}|^2 = 2|\int\Psi_m^* xy \Psi_n d^3r|^2 + \frac{1}{2}
|\int\Psi_m^* (x^2-y^2)\Psi_n d^3x|^2 . \label{dsqr}
\end{eqnarray}\\

To compute the mean transition rate for GW's incident from all
directions we simply  average $|D_{jk}e_{jk}|^2$ over the sphere.
The following is due to Dyson \cite{Dyson05}.  Since electron
wave functions do not depend on the direction of the
gravitational field, we can introduce two orthogonal
unit vectors $\hat\lambda, \hat\mu$ and rewrite Eq. (\ref{dsqr})
as
\begin{eqnarray}
|D_{jk}e_{jk}|^2 = 2|\int\Psi_m^* {\bf \hat\lambda \cdot r \mu
\cdot r } \Psi_n d^3r|^2 + \frac{1}{2}|\int\Psi_m^* ({\bf
\hat\lambda\cdot r \hat\lambda \cdot r - \hat\mu \cdot r \hat\mu
\cdot r }\Psi_n d^3x|^2 .\nonumber \label{dsqr2}
\end{eqnarray}
The average of this expression over all directions is then
\begin{equation}
\begin{array}{lll}
<|D_{jk}e_{jk}|^2> &=&  \frac{1}{4\pi}\int d\Omega \, [2
\hat\lambda_i\hat\mu_jD_{ij}\hat\lambda_k \hat\mu_l D_{kl}^* +
\hat\lambda_i\hat\lambda_jD_{ij}\hat\lambda_k\hat\lambda_lD_{kl}^*  \vspace{3mm}\nonumber\\
& -&
\frac{1}{2}\hat\lambda_i\hat\lambda_jD_{ij}\hat\mu_k\hat\mu_lD_{kl}^*
-
\frac{1}{2}\hat\mu_k\hat\mu_lD_{kl}\hat\lambda_i\hat\lambda_jD_{ij}^*],
\label{dsqr3} \vspace{2mm}
\end{array}
\end{equation}
where repeated indices are summed and
\begin{eqnarray}
<\hat\lambda_i\hat\lambda_jD_{ij}\hat\lambda_k\hat\lambda_lD_{kl}^*>
\, =\,
<\hat\mu_i\hat\mu_jD_{ij}\hat\mu_k\hat\mu_lD_{kl}^*>.\nonumber
\end{eqnarray}
Because ${\bf \hat\mu}$ and ${\bf \hat\lambda}$ are orthogonal, we can
eliminate ${\bf \hat\mu}$ by the following trick: Pick a direction for
${\bf \hat\lambda}$, say ${\bf \hat\lambda} = {\bf \hat k}$.  Then in the
usual spherical coordinates ${\bf \hat\mu} = {\bf  \hat i } cos\phi +
{{\bf \hat j }} sin \phi$.  The average $<\hat\mu_k\hat\mu_l>$
over the unit circle in the plane perpendicular to ${\bf \hat\lambda}$
can be seen to be $<\hat\mu_k\hat\mu_l> = 1/2\delta_{kl} -
1/2\hat\lambda_k\hat\lambda_l$. Since this is a tensor equation it
is true in any coordinate system and, hence, for any direction
${\bf \hat\lambda}$. Inserting this expression into Eq. (\ref{dsqr3})
and making use of the identity
\begin{eqnarray}
<\hat\lambda_i\hat\lambda_j\hat\lambda_k\hat\lambda_l> \equiv
\frac{1}{4\pi}\int d\Omega\,
\hat\lambda_i\hat\lambda_j\hat\lambda_k\hat\lambda_l =
\frac{1}{15}(\delta_{ij}\delta_{kl}+ \delta_{ik}\delta_{jl}+
\delta_{il}\delta_{jk}) \nonumber
\end{eqnarray}
 yields
\begin{eqnarray}
<|D_{jk}e_{jk}|^2>  = \frac{2}{5}(D_{jk}D_{jk}^* -
\frac{1}{3}D_{jj}D_{kk}^*).\label{finalavg}
\end{eqnarray}

It is this average that must be used in Eq. (\ref{abs2}) to
compute $\Gamma$. For the transition from the 1s to the 3d2 state
of hydrogen, the appropriate values of  of $D_{jk}$ are, from Eqs.
(\ref{wavefcns1}) and (\ref{Dqm}),
\begin{eqnarray}
D_{zz} &=& D_{xz} =  D_{yz} = 0\, ; \nonumber \\
D_{xx} &=& - D_{yy} =  -iD_{xy} = \frac{3^4 a^2}{2^8}\nonumber .
\end{eqnarray}
Substituting these values into Eq.( \ref{finalavg}) yields
\begin{eqnarray}
<|D_{jk}e_{jk}|^2>  = \frac{8}{5}|D_{xx}|^2 =
\frac{3^8 a^4}{5 \times 2^{13}}.  \label{dfin}
\end{eqnarray}
Finally, the absorption rate for the 1s to 3d2 transition follows
from Eq. (\ref{abs3}):
\begin{eqnarray}
\Gamma = \frac{3^8 \pi^2 Gm_e^2 a^4 \omega_{mn}^2 I(\omega_{mn})}
{5 \times 2^{13}\hbar^2 c^3} \label{abs4}
\end{eqnarray}
This expression also gives the stimulated emission rate between
the 3d2 and 1s hydrogenic states.\\

To check the consistency of the absorption rate with the
spontaneous emission rate calculated in \S\ref{spont}, we assume
that gravitational radiation within a cavity is in thermal
equilibrium with emitters and absorbers in the walls at
temperature $T$. (While this {\em detailed balance} argument is
``valid'' in a certain sense, it is straightforward to show that
no such cavity can in principle be constructed for gravitational
waves.) Equating the radiation absorbed per unit time with that
emitted per unit time we have
 \[
N_{3d2}(\Gamma_{sp} + \Gamma_{st}) = N_{1s}\Gamma_{ab},
 \]
where $\Gamma_{sp}$, $\Gamma_{st}$, and $\Gamma_{ab}$ are the
transition rates for spontaneous emission, stimulated emission,
and absorption, $N_{3d2}$ is the number of atoms in the 3d2 state,
and $N_{1s}$ is the number atoms in the 1s state.  Following
Einstein, we expect $N_{3d2}/N_{1s}= e^{-\hbar \omega_{mn}/kT}$.
 Now, substitute Eq. (\ref{abs4}) and the {\em spontaneous}
emission rate from Eq. (\ref{Gamma})  into the above expression.
Solving for $I(\omega_{mn})$ yields
\begin{eqnarray}
I(\omega_{mn}) = \frac{\hbar \omega_{mn}^3}{\pi^2 c^2} (e^{\hbar
\omega_{mn}/kT} -1)^{-1} \nonumber
\end{eqnarray}
which is consistent with the intensity of black body radiation.
This indicates that the ratio $\Gamma_{sp}/\Gamma_{ab}$ is correct
and  that the result yielded by the semi-classical approach in
\S\ref{spont} is as valid as Eq. (\ref{abs4}).

The absorption rate Eq. (\ref{abs4}) can be expressed in terms of an
integrated cross section,
\begin{equation}
 \Gamma = \frac{1}{\hbar \omega}\int \sigma(\omega) I(\omega)
d\omega.
\end{equation}
Clearly the absorption cross section is sharply peaked near
$\omega_{mn}$, and so $\int \sigma(\omega) d\omega = \hbar \omega
\Gamma/I(\omega_{mn})$.  If we define an average cross section as
$<\sigma> = \int \sigma(\omega_{mn}) d\omega / \omega_{mn}$, then
from Eq. (\ref{abs4}):
\begin{eqnarray}
<\sigma> = \frac{3^8 \pi^2 Gm_e^2 a^4 \omega_{mn}^2} {5 \times
2^{13}\hbar c^3} . \label{sigma}\nonumber
\end{eqnarray}
For the transition between the 1s and 3d2 state $\omega_{mn} =
\frac{4e^2}{9\hbar a}$, yielding
\begin{eqnarray}
\sigma_{abs} = \frac{3^4 \pi^2}{5 \times 2^9} \frac{G\hbar}{c^3} =
0.31 \ell_{pl}^2 .\label{sigma2}
\end{eqnarray}
Here, $\ell_{pl}\sim 10^{-33} cm$ is the Planck length.
Surprisingly, all the physical constants associated with the
hydrogen atom have disappeared from the cross section, leaving
only the square of the Planck length and a numerical constant of
order unity. We return to this important point in \S\ref{ion}.

The absorption rate (\ref{abs4}) was for unpolarized GWs averaged
over all incoming directions.  For completeness, the absorption
rate for  one polarization of a gravitational wave incident in the
$\theta$, $\phi$ direction is readily shown to be
\begin{eqnarray}
\Gamma = \frac{3^8 \pi^2 Gm_e^2 a^4 \omega_{mn}^2 I(\omega_{mn})}
{2^{16}\hbar^2 c^3}\left[(1+cos^2\theta)^2cos^22\phi +
4cos^2\theta sin^22\phi\right]. \nonumber
\end{eqnarray}
The rate for the other polarization is obtained by interchanging
the $sin^2 2 \phi$ and $cos^2 2 \phi$.

\section{Field Theoretic Calculation of Spontaneous Graviton Emission}
\setcounter{equation}{0} \label{detection2}

As an independent check of Eq. (\ref{Gamma}) we next compute the
transition rate via a field theoretic approach, that is, in terms
of gravitons. This  also allows a comparison with the result of
Lightman et al., who
 in problem 18.18 of their well-known {\it Problem Book in General Relativity
and Gravitation} \cite{Lightman75} also used  field theoretic
methods  to compute the 3d0 to 1s transition rate in hydrogen.

To quantize gravitational waves in the linearized theory we follow
the standard procedure of decomposing  the metric perturbations
into plane waves:
\begin{eqnarray}
h_{jk} = \frac{1}{\sqrt{V}} \sum_{{\bf k},\alpha} h_{{\bf
k},\alpha} e^{{\bf k},\alpha}_{jk}e^{i({\bf k \cdot r}-\omega t)}
+ c.c. .\label{waves}
\end{eqnarray}
Here, $h_{{\bf k},\alpha}$ are the Fourier amplitudes, $\alpha$
indicates the polarization, $e_{jk}$ is the polarization tensor,
and box normalization with volume $V$ is assumed.  The energy
density of GWs corresponding to the flux $\cal F$ given by Eq.
(\ref{flux}) is
 \[
dE/dV = \frac{c^2}{32 \pi G} \langle h^{TT}_{jk,0} h^{TT}_{jk,0}
\rangle.
 \] Substituting Eq. (\ref{waves}) into this expression
and integrating over all space gives the total energy in the GWs.
Noting that
 \[
\langle \int e^{i({\bf k \cdot r}-\omega t)}
e^{i({\bf k' \cdot r}-\omega' t)} d^3r \rangle = 0
 \] and
 \[
\langle \int e^{i({\bf k \cdot r}-\omega t)}
e^{-i({\bf k' \cdot r}-\omega' t)} d^3r \rangle = V
\delta_{\bf k k'}\delta_{\omega\omega'}
 \]
then
\begin{eqnarray}
E = \frac{c^2\omega^2}{16\pi G} \sum_{{\bf k},\alpha, \alpha'}
e^{{\bf k},\alpha}_{jk}e^{{\bf k},\alpha'}_{jk}
h_{{\bf k},\alpha} h^*_{{\bf k},\alpha'}.
\nonumber
\end{eqnarray}
For the normalization $|e_{kj}| = 1/\sqrt 2$, we have $\sum_{jk}
e^{{\bf k},\alpha}_{jk}e^{{\bf k},\alpha'}_{jk} =
\delta_{\alpha\alpha'}$, and
\begin{eqnarray}
E = \frac{c^2\omega^2}{16 \pi G} \sum_{{\bf k},\alpha} |h_{{\bf k},\alpha}|^2 .
\nonumber
\end{eqnarray}

Next identify $h_{{\bf k},\alpha}^*$ and $h_{{\bf k},\alpha}$ with
the raising and lowering operators, $a^\dag_{{\bf k},\alpha}$ and
$a_{{\bf k},\alpha}$, such that the classical energy and quantum
Hamiltonian agree with one another, i.e.,
\begin{eqnarray}
E = H = \frac{1}{2}\sum_{{\bf k},\alpha}\hbar\omega
(a^\dag_{{\bf k},\alpha}a_{{\bf k},\alpha} + a_{{\bf k},\alpha}a^\dag_{{\bf k},\alpha})
= \sum_{{\bf k},\alpha}(N_{{\bf k},\alpha} +\frac{1}{2})\hbar\omega
\nonumber
\end{eqnarray}
where $N_{{\bf k},\alpha} = a^\dag_{{\bf k},\alpha}a_{{\bf k},\alpha}$
is the number operator. One can always reset the zero of the energy scale so that
$H = \sum_{{\bf k},\alpha} \hbar\omega N_{{\bf k},\alpha}$.  The only way that $E = H$
is if we make the identifications
\begin{eqnarray}
h_{{\bf k},\alpha} \rightarrow \frac{1}{c}\sqrt{\frac{16\pi
G\hbar}{\omega}}a_{{\bf k},\alpha} \nonumber
\end{eqnarray}
and
\begin{eqnarray}
h_{{\bf k},\alpha}^* \rightarrow \frac{1}{c}\sqrt{\frac{16\pi
G\hbar}{\omega}}a^\dag_{{\bf k},\alpha} . \nonumber
\end{eqnarray}
Then
\begin{eqnarray}
h_{jk} \rightarrow \frac{1}{c\sqrt{V}}\sum_{{\bf
k},\alpha}\sqrt{\frac{16\pi G\hbar}{\omega}} \left[a_{{\bf
k},\alpha}e^{{\bf k},\alpha}_{jk}\exp{i({\bf k \cdot r}-\omega t)}
+ a^\dag_{{\bf k},\alpha}e^{{\bf k},\alpha}_{jk}\exp{-i({\bf k
\cdot r}-\omega t)}\right]. \nonumber\\
\label{operator}
\end{eqnarray}

We now consider the transition rate between two hydrogen states
that involves the emission of a single graviton.  According to the
{\it golden rule} the transition rate per solid angle is given by
\begin{eqnarray}
\frac{d\Gamma}{d\Omega} = \frac{2\pi}{\hbar}|\langle1|\langle
f|H|i\rangle|0\rangle|^2 \rho .\label{rate}
\end{eqnarray}
Here, $|i\rangle|0\rangle$ is the initial electron and graviton
state (with no gravitons), $|f\rangle|1\rangle$ is the final
electron and graviton state (with 1 graviton), and
\begin{eqnarray}
\rho = \frac{V\omega^2}{(2\pi)^3\hbar c^3}  \nonumber
\end{eqnarray}
is the graviton energy density of states per solid angle, again
box normalized. From Eqs. (\ref{norm}) and (\ref{hamiltonian1}), the
interaction Hamiltonian (in the local inertial frame) can be
written
\begin{eqnarray}
H = \frac{m_e \omega^2}{4} h_{jk}x^jx^k \label{hamilton}
\end{eqnarray}
where $h_{jk}$ is expressed in the TT gauge.  In order to compare
directly with the results of Lightman et al.(\cite{Lightman75}) we compute the
spontaneous transition between the 3d0 and 1s states of hydrogen,
although as pointed out earlier, the transition rates from all the
3dm states are identical.  Substituting Eq. (\ref{operator}) into
Eq. (\ref{hamilton}) the transition matrix element becomes
\begin{eqnarray}
\langle 1|\langle 1s|H|3d0\rangle|0\rangle &= &
\frac{m_e\omega^2}{4c} \sqrt{\frac{16\pi G\hbar}{V\omega}} \int
d^3r \Psi_{3d0}^\ast e_{jk} x^j x^k \Psi_{1s}\label{matrix}
\end{eqnarray}
where $\omega = (E_{3d0} - E_{1s})/\hbar$.  Substituting Eq.
(\ref{matrix}) into Eq. (\ref{rate}) yields
\begin{eqnarray}
\frac{d\Gamma}{d\Omega} = \frac{m_e^2\omega^5 G}{4\pi \hbar c^5}|D_{jk}e_{jk}|^2
\nonumber
\end{eqnarray}
where $D_{ij}$ is defined in Eq. (\ref{Dqm}).  As before, $
|D_{jk}e_{jk}|^2$ should be averaged over all directions. With
 Eq. (\ref{finalavg}) we then have:
\begin{eqnarray}
\Gamma = \frac{2m_e^2\omega^5 G}{5\hbar c^5}(D_{jk}D_{jk}^* -\frac{1}{3}D_{jj}D_{kk}^*) .
\nonumber
\end{eqnarray}
For the 3d0 and 1s hydrogen wavefunctions we find that $D_{xx} =
D_{yy} = -\frac{1}{2}D_{zz} = -a^2\sqrt{3^7/2^{15}}$ and,
therefore,
\begin{eqnarray}
\Gamma = \frac{3^8 Gm_e^2 a^4 \omega^5}{5 \times 2^{13}\hbar c^5},  \label{Gamma2}
\end{eqnarray}
which is precisely the same as the semi-classical result in Eq.
(\ref{Gamma}).  It is this agreement that provides the strongest
confirmation of the semi-classical treatment.

The {\it Problem Book} authors obtained a result that  is larger
than this one by about an order of magnitude.  In part the
disagreement is due to numerical and normalization errors.
Nevertheless, when these mistakes are corrected their
result still differs from
 Eq. (\ref{Gamma2}). The underlying reason can be traced to the fact
 that whereas our calculation was carried out in  the local inertial frame, Lightman
et al.\cite{Lightman75} worked entirely within the TT gauge.  One
can recover the proper result in the TT gauge; however, much care
must be taken to correctly interpret the interaction Hamiltonian
in that gauge, as will become apparent in the following section.

\section{Gauge Properties of the Interaction Hamiltonian}
\setcounter{equation}{0} \label{hamiltonian}

In his pioneering work on gravitational wave detection Weber
\cite{Weber61} used the equation of geodesic deviation to deduce
the gravitational force density of weak gravitational fields
acting on non-relativistic matter.  In a local inertial frame the
gravitational force density $f_g^j$ is
\begin{eqnarray}
f_g^j = -\rho R^j_{0k0} x^k , \label{geodev}
\end{eqnarray}
where $\rho$ is the mass density of the detector, and $R^j_{0k0}$
are components of the Riemann curvature tensor. Then the equation
of motion of the detector is $\rho \partial ^2x^j/\partial t^2 = f^j + f_g^j$
where $f^j$ is the total classical (i.e.,
non-relativistic) force density on the detector mass elements.   This
relation implicitly assumes that the gravitational wave has
negligible effect on the physics of the detector, which is a
reasonable assumption because the LIF is the most nearly
Minkowskian frame. Since gravity couples to the
mass-energy density of matter, one  expects that the gravitational
modification of the classical forces in the detector will be
proportional to the binding energy density of the system. In fact,
it can be shown that this is the case \cite{Boughn75}. While Eq.
(\ref{geodev}) strictly holds true only in a LIF, $R^j_{0k0}$ is
already first order in $h_{\mu \nu}$, and so the right hand side
of the equation is, to first order, invariant under infinitesimal
coordinate, or gauge, transformations.  It is in this sense that
the expression for the gravitational force in Eq. (\ref{geodev})
is gauge invariant.

As an alternative formulation to Weber's, Dyson \cite{Dyson69}
introduced the interaction Lagrangian density, $\mathcal{L} =
\frac{1}{2} h_{\mu \nu} T^{\mu \nu}$, which we used in
\S\ref{balance} and \S\ref{detection2} above.  It follows directly
from the general definition of the stress-energy tensor as the
functional derivative of the matter action, $I_m$, with respect to
$g_{\mu\nu}$ \cite{Weinberg72}:
\begin{eqnarray}
\delta I_m = \frac{1}{2} \int d^4x \sqrt{g(x^\mu)} T^{\mu
\nu}(x^\mu)\delta g_{\mu \nu}(x^\mu). \nonumber
\end{eqnarray}
In this expression $g(x^\mu)$ is the determinant of $-g_{\mu
\nu}(x^\mu)$ and $g_{\mu \nu}$ is to be considered an external
field rather than a dynamical variable.  For any metric that is
close to Minkowskian, i.e., $g_{\mu \nu} = \eta_{\mu \nu} + h_{\mu
\nu}$ and $\sqrt{g(x^\mu)} \approx 1$, one can easily see that
 the action differs from that of a free particle by
$\frac{1}{2} \int d^4x T^{\mu \nu}(x^\mu)h_{\mu \nu}(x^\mu)$.
Therefore, the interaction Lagrangian density is given by
\begin{eqnarray}
\mathcal{L}_I = \frac{1}{2} h_{\mu\nu}T^{\mu\nu} . \label{lagrangian}
\end{eqnarray}

The Euler-Lagrange equations following from a Lagrangian with this
interaction term are the same as implied by the equation of geodesic
deviation  only if $h_{\mu \nu}$ is expressed in the LIF and
$T^{\mu \nu}$ is the classical stress-energy tensor of the system.
The reason is that the Lagrangian  (\ref{lagrangian}) is not
invariant under infinitesimal coordinate (gauge) transformations
$x'_\mu = x_\mu + \xi_\mu$, even though the action from which it
is derived is a scalar quantity.  Of course, this doesn't mean
that the  computed  motions of particles depend on a particular
gauge. The $T^{\mu\nu}$ in Eq. (\ref{lagrangian}) includes the
effects of the classical, non-gravitational forces acting on the
particle. In non-LIF gauges, e.g., the TT gauge, these forces can
be significantly modified by the gravitational wave and the
modifications must be taken into account in calculating particle
motions.

As a specific example, consider a system of particles held
together by electromagnetic forces.  In principle, the action must
therefore include the electromagnetic stress-energy, which is
coupled to the gravitational field.  In the LIF gauge
gravitationally induced electromagnetic forces are smaller than
the tidal forces and can be neglected. On the other hand, in  a
non-LIF, the solution to the Einstein-Maxwell equations include
gravitationally induced electromagnetic forces that are comparable
to the tidal effects and, therefore, they must be taken into
account.  (A simple example of such a system was treated by Boughn
\cite{Boughn75}.)

Dyson's treatment of elastic systems in the presence of
gravitational waves \cite{Dyson69} is based on the interaction
Lagrangian (\ref{lagrangian}) as expressed in the TT gauge, which
does not constitute a LIF.  Gravitationally induced modifications
to the internal stresses must therefore be taken into account.
Dyson does not do this, however, and concludes that ``the response
[of an elastic solid] depends on irregularities in the shear-wave
modulus, and is strongest at free surfaces.''  As a consequence,
Dyson's analysis implies that a self-gravitating, compressible perfect fluid
(a system without shear), should not interact with a gravitational
wave. But in fact the sun is a reasonable approximation to such a
system at periods comparable to sound travel times across the sun
and yet does couple to incident gravitational waves
\cite{boughn84,boughn85}.

Similarly, in computing their transition rate in hydrogen, the
{\it Problem Book} authors \cite{Lightman75}
worked in the TT gauge but did not include
electromagnetic stresses in the interaction Lagrangian. Recall
that in special relativity, the Minkowski metric can be written
$d\tau^2 = -\eta_{\mu\nu}dx^\mu dx^\nu = dt^2(1-v^2)$.  This leads
to the free-particle Lagrangian $L = -m\sqrt{ 1-v^2}$. For a
charged particle interacting with both an electromagnetic and a
gravitational field we replace $\eta_{\mu\nu}$ by $g_{\mu\nu} =
\eta_{\mu\nu} + h_{\mu\nu}$ and add the usual electromagnetic
interaction term, giving
\begin{equation}
L = -m{\sqrt{ 1-\eta_{jk}v^jv^k - h_{jk}v^j v^k}} - q\Phi + qv^jA_j,
 \label{lagrangianf}
\end{equation}
where $\Phi$ and ${\bf A}$ are the  electromagnetic scalar and
vector potentials and we assume the $h_{\mu 0} = 0$ condition of
the TT gauge.  For non-relativistic systems Eq. (\ref{lagrangianf}) 
can be expanded to lowest order in $v^j$,
\begin{equation}
L = -m +\frac{1}{2}m (\eta_{jk}v^jv^k + h_{jk}v^j v^k) - q\Phi + qv^jA_j.
 \label{lagrangianff}
\end{equation}
Note that the $h_{jk}$ term in this expression is just the interaction 
Lagrangian of Eq. (\ref{lagrangian}).  The Hamiltonian is formed through 
the standard prescription  $H \equiv \pi_\alpha \dot q^\alpha - L$, where
$\pi_\alpha \equiv \partial L/\partial {\dot q^\alpha}$ and $q^\alpha$
are the generalized momenta and velocities.  Working to first order in
$h_{jk}$ the canonical momenta are
\[
\pi_j = m(\eta_{jk}v^k + h_{jk}v^k) + q A_j.
        \nonumber
\]
To first order, the inverse of $(\eta_{jk} + h_{jk})$ is $(\eta^{jk} - h^{jk})$ 
where $h^{jk} \equiv h_{jk}$.  Then solving for $v^j$ gives
\[
v^j = \frac{1}{m}(\eta^{jk} - h^{jk})(\pi_k - q A_k)
        \nonumber
\]
and the Hamiltonian becomes
\begin{equation}
H = \frac{(\eta^{jk} - h^{jk})(\pi_j - qA_j)(\pi_k - qA_k)}{2m} + m +q\Phi.
\label{hamiltonianp}
\end{equation}
The constant $m$ is not physically relevant and, as one sees, the Hamiltonian
(\ref{hamiltonianp}) follows the usual minimal substitution rule,
 ${\bf \pi} \rightarrow {\bf \pi} - q {\bf A}$.   To implement
 Eq. (\ref{hamiltonianp}),
$\pi_j$ is identified with the quantum mechanical operator $-i
\hbar
\partial /\partial x^j$ and $h_{jk}$ is identified with the graviton
raising and lower operators according to Eq. \ref{operator}.  For
nonrelativistic systems, magnetic fields are much smaller than
electric fields and so, in the Coulomb gauge, the vector potential {\bf A} can
be ignored.  In this case the interaction part of the Hamiltonian operator
becomes
\begin{equation}
H_I = \frac{\hbar^2}{2m} h_{jk} \frac{\partial}{\partial x_j}
\frac{\partial}{\partial x_k}. \label{hamiltonianp2}
\end{equation}
When this expression is substituted into the matrix element of Eq.
(\ref{rate}) the result is a transition rate  $16/25$ times that
of Eq. (\ref{Gamma2}) and is the rate presented in the {\it Problem
Book}, modulo numerical and normalization errors.\\

This factor can be accounted for by including the purely electromagnetic
part of the Lagrangian density
\begin{equation}
{\cal L}_{EM} = -\frac{1}{16\pi}F_{\mu\nu} F^{\mu\nu}
=  -\frac{1}{16\pi}g^{\alpha\mu}g^{\beta\nu} F_{\alpha\beta}F_{\mu\nu}
\label{lagrangianem}
\end{equation}
where $F_{\mu\nu} = A_{\nu,\mu} - A_{\mu,\nu}$ is the usual electromagnetic 
field tensor \cite{Jackson75}. To lowest order in $h^{jk}$  this becomes
\begin{equation}
{\cal L}_{EM} =  -\frac{1}{16\pi}\eta^{\alpha\mu}\eta^{\beta\nu}F_{\alpha\beta}F_{\mu\nu}
+\frac{1}{8\pi}h^{jk}\eta^{\mu\alpha}F_{\mu j}F_{\alpha k}. 
\label{lagrangianem2}
\end{equation}
The purely electromagnetic part of the Hamiltonian density then has the form
\begin{equation}
{\cal H}_{EM} = (\partial {\cal L}_{EM}/\partial A_{j,0})A_{j,0} - {\cal L}_{EM}.
\nonumber
\end{equation}
It is straightforward to show that to first order in $h^{jk}$ the total Hamiltonian 
becomes (see, for example, Chapter 57 of Schiff \cite{Schiff49})
\begin{equation}
H = \frac{(\eta^{jk} - h^{jk})(\pi_j - qA_j)(\pi_k - qA_k)}{2m} +
\frac{1}{8\pi} \int (|{\bf E}|^2 d^3x + h^{jk}E_jE_k) d^3x
\label{hamiltonianp3}
\end{equation}
where we have again invoked the Coulomb gauge and ignored terms involving the
magnetic field so the electric field is given by $E_j = A_{0,j}$.  
Finally, the interaction part of the Hamiltonian can be written as
\begin{equation}
H_I = \frac{\hbar^2}{2m} h_{jk} \frac{\partial}{\partial x_j}
\frac{\partial}{\partial x_k} +\frac{1}{8 \pi} \int  h_{jk}E^jE^k d^3x 
\label{hamiltonianp4}
\end{equation}
since $h^{jk}=h_{jk}$ and to lowest order in $h_{jk}$ there is no distinction
between upper and lower indices in the other fields.

In this expression $E^j$ is the sum of electric fields of the
proton and the electron, $E^j_p +E^j_e$, so the integrand of Eq.
(\ref{hamiltonianp4}) is $(E^j_p + E^j_e)(E^k_p + E^k_e)$.  The
integrals of $E^j_pE^k_p$ and $E^j_eE^k_e$ over all space are
constants, independent of the location of the particles.
Therefore, the contribution of these terms to the matrix element
(see Eq. \ref{rate}) between the $1s$ and $3d0$ states vanishes.
The only electromagnetic terms in the interaction Hamiltonian that
contribute to the transition rate are
\begin{equation}
H_{EM,I} = \frac{1}{8 \pi}h_{jk} \int (E^j_pE^k_e + E^k_pE^j_e) d^3x 
\label{hamiltonianem3}
\end{equation}
where we again assume that $h_{jk}({\bf r})$ is approximately constant 
in the region of significant electric fields..
Consider the first term
\begin{eqnarray}
E_p^jE_e^k = -e^2 \frac{x^j (x^k-x_e^k)}{r^3 |{\bf r - r}_e|^3} =
e^2 \frac{x^j}{r^3}\nabla^k \frac{1}{|{\bf r - r}_e|}.
\label{efield}
\end{eqnarray}
Here $e$ is the electron charge, ${\bf r}$ is the field point,
${\bf r}_e$ is the electron
coordinate, and the proton is assumed to be located at ${\bf r}
=0$. Via the addition theorem for spherical harmonics,  $ |{\bf r
- r}_e|^{-1}$ can be expressed as \cite{Jackson75},
\begin{eqnarray}
\frac{1}{|{\bf r - r}_e|} = 4\pi \sum_{\ell = 0}^\infty \sum_{m =
-\ell}^\ell \frac{1}{2\ell + 1} \frac{r_<}{r_>^{\ell + 1}}
Y^*_{\ell m}(\theta',\phi') Y_{\ell m}(\theta,\phi). \label{ylm}
\nonumber
\end{eqnarray}
Noting that the matrix element between the 1s and 3d0 states will
pick out only the $\ell = 2, m=0$ term, we are thus left with
\begin{eqnarray}
\frac{1}{|{\bf r - r}_e|} = \frac{4\pi r_<^2}{5
r_>^3}Y^*_{20}(\theta,\phi) Y_{20}(\theta_e,\phi_e), \label{y20}
\end{eqnarray}
where $r_<$ ($r_>$) is the smaller (larger) of $r$ and $r_e$.
Substituting Eq. (\ref{y20}) into Eq. (\ref{efield}) and
integrating over all space yields, after a tedious calculation,
\begin{eqnarray}
\int d^3x E_p^xE_e^x = \int d^3x E_p^yE_e^y = -\frac{1}{2} \int
d^3x E_p^zE_e^z = -\frac{\pi}{3} \frac{e^2(3\cos^2\theta_e -
1)}{r_e}.  \label{ee}
\end{eqnarray}
The integrals vanish for $j \neq k$ and, therefore, the two terms in
Eq. (\ref{hamiltonianem3}) are equal.  Substituting Eq. (\ref{ee}) into
Eq.(\ref{hamiltonianem3}) and adding the result to the electron Hamlitonian
(see Eq. \ref{hamiltonianp2}) yields the interaction part of the Hamiltonian.
Another tedious computation via Eq. (\ref{rate}) gives, finally, a
transition rate of
\begin{eqnarray}
\Gamma = P/\hbar\omega = \frac{3^8 Gm_e^2 a^4 \omega^5}{5 \times
2^{13}\hbar c^5} = \frac{\alpha^6G{m_e}^3c}{360\hbar^2},
\label{Gamma3}
\end{eqnarray}
which is precisely the same as the LIF field theoretic and
semi-classical results (\ref{Gamma2}) and (\ref{Gamma}).

From this analysis we conclude that one can work with an
interaction Hamiltonian in any gauge as long as all the revelent
interactions are included in the stress-energy tensor. However, it
is also clear that in this case  working in the LIF leads to a
much simpler analysis. For classical systems, it is undoubtedly
also possible to work in non-LIF gauges; however, in the case of
elastic media considered by Dyson \cite{Dyson69}, for example,
the best way to do this is not immediately apparent.

\section{Graviton Ionization cross section of Hydrogen}
\setcounter{equation}{0} \label{ion}

In RB, an important step was to calculate the ionization cross
section for hydrogen.  As we saw in \S\ref{balance}, the graviton
absorption cross section was $\sigma \sim \ell_{Pl}^2$. That all
dimensions other than the Planck length drop out of the cross
section is at first surprising, but can be understood simply as
follows:  The classical cross section for a system of mass $m$,
size $\ell$, and frequency $\omega$ is roughly \cite{Misner73,
RB06} $\sigma \sim Gm\ell^2 \omega/c^3$. Assuming the
Nicholson-Bohr quantization condition, the angular momentum for
such a system near its ground state is $L \sim m\ell^2 \omega \sim
\hbar$.  Thus
\begin{eqnarray}
\sigma \sim \frac{G\hbar}{c^3} = \ell_{Pl}^2 \approx 10^{-66}
cm^2,  \nonumber
\end{eqnarray}
and we see that the Planck-length-squared cross section is solely
a result angular momentum quantization.

Because in RB we decided on the gravitational analogy of the
photoelectric effect as a method for detecting gravitons, it was
necessary to compute the gravito-ionization cross section for
hydrogen in the ground state.  One expects it to be similar in
magnitude to the above; however the ordinary photoionization cross
section does have large numerical factors ``of order unity" and is
also strongly dependent on photon energy.  Therefore, in this
section we compute the ionization cross section for gravitons with
energies in the range $13.6 eV \ll E \ll 2.5 \times 10^4 eV$,
energies for which the non-relativistic Born approximation is
appropriate.  In fact, as detailed in RB, there are many strong
astrophysical sources of gravitons in this energy range.

As in \S\ref{balance}, we take the interaction Hamiltonian in a
LIF to be
\begin{eqnarray}
H = \frac{1}{4}m_e \omega^2 h x^j x^k e_{jk} \exp{i(k_lx^l-\omega t)} + c.c. \label{hamiltonian2}
\end{eqnarray}
and compute the matrix element between an initial hydrogen ground state, $\Psi_i$,
and a plane wave final state (the Born approximation), $\Psi_f$, i.e.,
\begin{eqnarray}
\Psi_i =\frac{1}{\sqrt{\pi} a^{3/2}}e^{-r/a} &;&\Psi_f =
\frac{1}{L^{3/2}}e^{i{\bf k \cdot r}} \label{ionwavfcns}
\end{eqnarray}
where the plane wave is box normalized with dimension $L$. The
transition probability per unit time between these two states is
given by the {\it golden rule},
\begin{eqnarray}
\Gamma = \frac{2\pi}{\hbar} \rho(k) |\langle f|H|i\rangle|^2
\label{gold}
\end{eqnarray}
where $k$ is the wave number of the emerging electron and
$\rho(k)$ is the energy density of final states:
\begin{eqnarray}
\rho(k) = \frac{m_e k L^3}{2 \pi^2 \hbar^2} .\label{rhof}
\end{eqnarray}
As in \S\ref{balance}, we average $|\langle f|H|i \rangle|^2$
over all directions using Eq. (\ref{finalavg}).  For an incident
GW of amplitude $h$ this result is
\begin{eqnarray}
|\langle f|H|i \rangle|^2 = \frac{3 \cdot 2^{11} \pi}{5}
\frac{h^2\omega^4 m_e^2a^7 (a^4k^4)}{L^3 (1+a^2k^2)^8}. \nonumber
\end{eqnarray}
Eqs. (\ref{gold}) and (\ref{rhof}) then give
\begin{eqnarray}
\Gamma = \frac{3 \cdot 2^{10}}{5} \frac{h^2\omega^4 m_e^3a^{11}k^5}{\hbar^3 (1+a^2k^2)^8} .
\nonumber
\end{eqnarray}
By definition, the ionization cross section is $\sigma = \Gamma
\hbar \omega / \mathcal{F}$ where $\mathcal{F} = \frac{c^3
\omega^2 h^2}{8 \pi G} h^2$ is the GW flux from \S\ref{balance}.
Thus,
\begin{eqnarray}
\sigma = \frac{3 \cdot 2^{13}\pi}{5}\frac{G\omega^3m^3a^{11}k^5}
{c^3\hbar^2(1+a^2k^2)^8} .\label{ion1}
\end{eqnarray}
$\omega$ can be eliminated from this expression by using the
Einstein photoelectric relation (conservation of energy), which
requires that the incident graviton energy equal the sum of the
electron binding energy and the kinetic energy of the emerging
electron, or $\hbar \omega = e^2/2a + k^2/2m_e$. Hence,
\begin{eqnarray}
\sigma = \frac{3 \cdot 2^{10}\pi}{5}\frac{(ka)^5}{(1+k^2a^2)^5}
\frac{G\hbar}{c^3} . \label{ion2}
\end{eqnarray}
For $ak >> 1$ the dependence of this result on final electron
momentum is in fact identical to that of the ordinary
photoionization cross section. We also see that the ionization
cross section is, modulo a dimensionless factor, equal to the
Planck length squared; for energetic gravitions with $\hbar \omega
>> 13.6$eV, however, the dimensionless factor can be quite small.
Smolin\cite{Smolin85} used a field theoretic argument to estimate the 
ionization cross section of a bound electron and obtained a result that is 
also proportional to $\ell_{Pl}^2$.  However, the dependence on $ka$ is quite
different, with the cross section increasing rapidly for large $ka$ rather
than decreasing rapidly as indicated in Eq. (\ref{ion2}).

The differential ionization cross section for linearly polarized
gravitons is also not difficult to compute and we state it here
for completeness.  For one polarization we obtain
\begin{eqnarray}
d\sigma/d\Omega = 3^2 \cdot 2^5\frac{(ka)^5}{(1+k^2a^2)^5}\sin^2
2\phi \sin^4\theta \frac{G\hbar}{c^3}; \nonumber
\end{eqnarray}
letting $\sin^2 2\phi \rightarrow \cos^2 2\phi$ gives the other
polarization.

\section{Conclusion}
\setcounter{equation}{0} \label{discussion}

As discussed in the Introduction, the original motivation for this
paper was Dyson's conjecture that a single graviton could never be
detected in the real universe.  In RB we employed the above
gravito-electric cross section to show that if one is limited only
by the mass, energy content, and age of the universe, one can
design a highly idealized {\em gedanken} experiment that could
detect {\em some} gravitons in the lifetime of the universe.  As
soon as one begins to consider detector physics and background
noise, though, detecting even a single graviton becomes
impossible.  In that sense, Dyson's conjecture appears correct.

Although one might argue that the detailed calculations presented
here are not entirely necessary to address Dyson's conjecture, it
has become clear that the physics of gravitational wave-matter
interaction, in particular the gauge properties of the interaction
Hamiltonian, present enough subtleties to catch even experienced
practitioners off guard.  As this will undoubtedly happen from
time to time in the future, we feel  it is of some
importance to elucidate this matter as clearly as possible.\\

{\bf Acknowledgements}
We would like to thank Freeman Dyson for inspiring this work and
for generously making available his unpublished notes.  We also thank
Jim Peebles for useful discusssions and the physics department at
Princeton University where much of this work was carried out.

{\small

\end{document}